\documentclass[useAMS,usenatbib]{mn2e}
\usepackage[dvips]{graphicx}
\usepackage[english]{babel}
\usepackage[latin1]{inputenc}
\usepackage{mathrsfs}
\usepackage{latexsym}
\usepackage{amssymb}
\bibpunct{(}{)}{,}{a}{}{,}%

\newcommand{\ie}{\textit{i.e.}}

\newcommand{\vsini}{v \cdot \sin i}
\newcommand{\veq}{v_{\mathrm{eq}}}
\newcommand{\kms}{\mathrm{km.s}^{-1}}
\newcommand{\Req}{R_\mathrm{eq}}
\newcommand{\Rp}{R_\mathrm{pol}}
\newcommand{\Rps}{R_\mathrm{pol}^{\star}}
\newcommand{\Rs}{R^{\star}}
\newcommand{\dd}{\textrm{d}}
\newcommand{\vect}{\vec}
\newcommand{\grad}{\vect{\nabla}}
\newcommand{\lapl}{\Delta}
\renewcommand{\div}{\vect{\nabla} \cdot}
\renewcommand{\l}{\ell}

\newcommand{\cost}{\cos\theta}
\newcommand{\vlp}{\left[-\omega + m\Omega\right]}
\newcommand{\Ylm}{Y^m_{\ell}}

\title[On the effects of rotation on acoustic stellar pulsations]
      {On the effects of rotation on acoustic stellar pulsations:
       validity domains of perturbative methods and close frequency pairs}
\author[K. D. Burke, D. R. Reese, and M. J. Thompson]
       {K. D. Burke$^1$\thanks{E-mail: K.Burke@sheffield.ac.uk},
        D. R. Reese$^{1,2}$ and M. J. Thompson$^{1,3}$ \\
        $^1$Department of Applied Mathematics,
        University of Sheffield,
        Hicks Building,
        Hounsfield Road,
        S3 7RH,
        Sheffield, UK \\
        $^2$Observatoire de Paris, LESIA, CNRS UMR 8109,
        92195 Meudon, France \\
        $^3$High Altitude Observatory, National Center for
        Atmospheric Research, Boulder, CO 80307}
\begin{document}

\date{Accepted ???. Received ???; in original form ???}

\pagerange{\pageref{firstpage}--\pageref{lastpage}} \pubyear{2010}

\maketitle

\label{firstpage}

\begin{abstract}
Pulsation frequencies of acoustic modes are calculated for realistic rotating stellar models using
both a perturbative and a two-dimensional approach.  A comparison between the
two yields validity domains which are similar to those previously obtained in
\citet*{Reese2006} for polytropic models.  One can also construct validity
domains based on polynomial fits to the frequencies from the two-dimensional approach, and these
also turn out to be similar, thus further confirming the agreement between the
perturbative and two-dimensional approach at low rotation rates.  Furthermore,
as was previously shown in \citet*{Espinosa2004}, adjacent frequencies in
multiplets come close together, thus forming pairs.  This phenomena, exclusive
to two-dimensional calculations, is shown to be an unlikely explanation of the
close frequency pairs observed in $\delta$ Scuti stars.  A systematic search for
all close frequency pairs in the calculated spectrum is also carried out.  The
number of close frequency pairs is shown to agree with what is expected based on
a Poisson distribution, but does not match the number or distribution of close
pairs in stars like FG Vir. Furthermore, a lack of close frequency pairs appears
at low rotation rates, where frequency multiplets do not overlap.  Delta Scuti
stars currently reported as having close frequency pairs do not fall in this
interval.
\end{abstract}


\section{Introduction}

The new space missions, CoRoT and Kepler, are substantially increasing the
accuracy with which stellar pulsations are observed.  A look at CoRoT's and
Kepler's first asteroseismic results shows the progress that has been made in
the photometric detection of solar-like oscillations, as well as the substantial
increase in the number of detected modes in earlier type stars such as $\delta$
Scuti stars \citep{Michel2008, Kjeldsen2010}.  The high duty cycle and the long
observational runs (up to 5 months for CoRoT and several years for Kepler) are
key factors in this dramatic improvement.  As a result, there is a drive on the
theoretical side to achieve accurate computations of pulsation frequencies in
stellar models.  Although a fair amount of success has been achieved in slowly
rotating stars, much work is still needed before being able to accurately model
pulsation modes in rapidly rotating stars.  This is because at rapid rotation
rates, the centrifugal force causes stellar deformation thereby transforming
what used to be a one-dimensional eigenvalue problem into one that is
two-dimensional.  This is all the more problematic as many stars rotate rapidly.

In order to address the effects of rotation, two different methods have been
developed. The first is the perturbative method which starts from a non-rotating
star and then includes various corrections which model the effects of rotation. 
These corrections are generally expressed as a power series in terms of the
rotation rate which is assumed to be a small parameter.  Previous works include
\citet{Saio1981, Gough1990} and \citet{Dziembowski1992} for the second order,
and \citet*{Soufi1998} and \citet{Karami2005} for the third order.  The advantage
of this method is that it is computationally less expensive and mode
classification is more straightforward, thereby allowing an efficient study of
an entire spectrum of pulsation modes for a given model.  However this method is
only valid for slow rotation rates.  The other approach consists in solving
directly the two-dimensional eigenvalue system.  Previous works which focus on
p-modes include \citet*{Clement1981, Yoshida2001, Espinosa2004, Lignieres2006}
and \citet{Lovekin2008}.  Although computationally more expensive, this method
can handle pulsation modes at any rotation rate.  A natural question is
at what rotation rates does it becomes necessary to use a two-dimensional
approach rather than a perturbative approach.

A number of previous works have focused on answering the above question in the
case of acoustic modes. In \citet*{Reese2006}, validity domains were established
for third order perturbative methods.  The underlying stellar models were
polytropic and the perturbative calculations were, in fact, polynomial fits to
the non-perturbative\footnote{The term ``non-perturbative frequencies'' in this context means
frequencies based on a two-dimensional calculation, rather than frequencies in which the effects
of rotation have been removed.} frequencies.  Later on, \citet{Lovekin2008} did
comparisons using more realistic models.  Once more the perturbative
calculations were polynomial fits to the non-perturbative frequencies. Given the
larger error bars on the calculations, perturbative calculations were considered
to be valid up to higher rotation rates.  Other works include
\citet{Ouazzani2009} where non-perturbative frequencies were compared with
frequencies resulting, this time, from a perturbative analysis rather
than a polynomial fit.  Furthermore, the effects of avoided crossings were
included in the perturbative calculations and shown to improve the agreement
with two-dimensional calculations.   Finally, \citet{Suarez2010}
investigated the effects of perturbative and non-perturbative frequencies
of polytropic models on echelle diagrams and various frequency separations. They
showed that including avoided crossings in perturbative analysis becomes
necessary for correctly predicting the behaviour of the small frequency
separation, even at relatively low rotation rates.  Nonetheless, a systematic
study of the validity of perturbative methods for realistic models is still
lacking.  Given the accuracy of the new pulsation data coming from the CoRoT and
Kepler missions, such a study is needed in order to correctly interpret this
data.  Therefore, in the present work, we will compare two-dimensional
calculations with frequencies resulting from a second order perturbative
analysis and establish validity domains similar to those in \citet{Reese2006}. 
These calculations will be carried out for acoustic modes, using realistic
stellar models rather than polytropic ones.  Furthermore, we will discuss some
of the qualitative differences between the two types of calculations.

The next section deals with the stellar models.  This is then followed by a
description of the perturbative and two-dimensional pulsation calculations.
Section~\ref{sect:comparison} compares the results from the two approaches and
section~\ref{sect:frequency_pairs} deals with close frequency pairs, both within
multiplets, and for an entire spectrum.  Finally, a discussion concludes the
paper.

\section{Models}

Stellar models were created using the Aarhus stellar evolution code, ASTEC
\citep{astec}. The OPAL 1995 opacity tables \citep{iglesias96} were used in
conjunction with the Kurucz low temperature adjustments \citep{kurucz91}, and
the Eggleton, Faulkner and Flannery equation of state \citep*{Eggleton1973}. 
Evolution begins at zero-age main sequence (ZAMS) and continues to a
user-specified age. The initial mass, $M$, angular velocity $\Omega$ and
hydrogen and heavy element abundances $X$ and $Z$ are also set by the user.
The models calculated here are $1 M_{\odot}$ and $2 M_{\odot}$ ZAMS models with solar elemental
abundances and rotation rates ranging from $\Omega/\Omega_{K}=0-0.25$, where
$\Omega_{K}=\sqrt{GM/R^{3}}$ and $R$ is the stellar radius. A selection
of the models calculated is given in Table~\ref{tab:models}.

The effects of rotation were then included to second order in rotational
velocity, $\Omega$.  In particular, the stellar structure is no longer
spherically symmetric but oblate due to the centrifugal force.  We therefore
perform a transformation of the coordinate system  $(r,\theta,\phi) \rightarrow
(x,\theta,\phi)$, from a system of shells of constant radius, $r$, to shells of
constant pressure, chosen such that $x=R$ maps the surface of the star which can
now be treated as a surface in hydrostatic equilibrium. The new radial
coordinate $x$ takes the form
\begin{equation}
\label{eqn:x}
x = \left[1+h_{\Omega}(r)P_{2}(\cos{\theta})\right]r,
\end{equation}
where $h_{\Omega}$ is the transformation coefficient and $P_2 (\cos\theta) =
\frac{3\cos^2\theta-1}{2}$ the $\ell=2$ Legendre polynomial. 
This can be inverted to give, to order $\Omega^{2}$
\begin{equation}
r = \left[1-h_{\Omega}(x)P_{2}(\cos{\theta})\right]x.
\end{equation}
We define a function $u=h_{\Omega}x$ such that $u$ satisfies the following
second order differential equation:
\begin{equation}
\label{eqn:kugx}
\mathscr{K}u = \mathscr{G}(x),
\end{equation}
where 
\begin{eqnarray}
\label{eqn:k}
\mathscr{K} &\equiv& \frac{{\dd}^{2}}{{\dd}x^{2}}+\left(
\frac{8\pi x^{2}\rho_{0}}{Mq_{0}} - \frac{2}{x}\right) \frac{{\rm
d}}{{\dd}x} - \frac{4}{x^{2}}, \\
\label{eqn:g}
\mathscr{G}(x) &=& \frac{4\pi
x^{4}\rho_{0}}{GM^{2}q_{0}^{2}}\left(f_{r2} - \frac{{\dd}}{{\rm
d}x}(xf_{\theta 2})\right) + \frac{6x}{GMq_{0}} \\
\nonumber
&& - \frac{1}{GMq_{0}}\frac{{\dd}}{{\dd}x}\left(x^{2}\frac{{\rm
d}}{{\dd}x}(xf_{\theta 2})\right),
\end{eqnarray}
and $f_{r0}=-\frac{2}{3}\Omega^{2}x$, $f_{r2}=\frac{2}{3}\Omega^{2}x$, $f_{\theta 2}=\frac{1}{3}\Omega^{2}x$ and $q_{0}=m/M$ is the fractional mass interior to a radius $x$.
Boundary conditions for this system are
\begin{eqnarray*}
&& u \quad \mathrm{regular \quad as} \quad x\rightarrow 0 \\
&& u' + \frac{u}{x} = -\frac{x^{3}}{GM}\left(x\frac{\dd}{\dd x}(f_{\theta 2})+ 4f_{\theta 2}\right)\qquad \mathrm{at }\quad x=R.
\end{eqnarray*}
The surface boundary condition is found by matching the gravitational potential
$\Phi_{0}$ onto the vacuum potential for $x > R$.

This change in coordinates leads to a change in all variables with a radial
dependence such that e.g. $\rho_{0}(x)=\rho_{0}(r)+\rho_{\Omega}(r)P_2
(\cos\theta)$, where the subscript ``$0$'' represents the equilibrium
quantity, with similar expressions for $p_{0}(x)$ and $c_{0}(x)$.

\begin{table*}
\begin{minipage}[t]{\columnwidth}
\begin{center}
\caption{Model parameters.}
\label{tab:models}
\begin{tabular}{c c c c c c c c}
\hline
\hline
 $\mathbf{M/M_{\odot}}$ & \textbf{Age} & $\mathbf{\Omega/\Omega_{K}}$ & $\mathbf{\Omega}$ & $\mathbf{R/R_\odot}$ & $\mathbf{L/L_\odot}$ & $\mathbf{T_{\rm{eff}}}$ & $\mathbf{h_{\Omega}(R)}$ \\
& \textbf{(Gyr)} & & \textbf{(rad.s$\mathbf{^{-1}}$)} & & & \textbf{(K)} & \\
\hline 
1.00	&	0.00	&	0.0            &       0.00            &       0.89    &       0.69    &       5573.9  &       0.00 \\
1.00	&	0.00	&	2.0$\times 10^{-5}$    &       1.50$\times 10^{-8}$    &       0.89    &       0.69    &       5573.9  &       1.40$\times 10^{-10}$\\
1.00	&	0.00	&	2.0$\times 10^{-4}$    &       1.50$\times 10^{-7}$    &       0.89    &       0.69    &       5573.9  &       1.40$\times 10^{-8}$\\
1.00	&	0.00	&	2.0$\times 10^{-3}$    &       1.50$\times 10^{-6}$    &       0.89    &       0.69    &       5573.9  &       1.40$\times 10^{-6}$\\
1.00	&	0.00	&	2.0$\times 10^{-2}$    &       1.50$\times 10^{-5}$    &       0.89    &       0.69    &       5573.5  &       1.40$\times 10^{-4}$\\
1.00	&	0.00	&	2.0$\times 10^{-1}$    &       1.49$\times 10^{-4}$    &       0.89    &       0.67    &       5534.6  &       1.44$\times 10^{-2}$\\
\hline
2.00	&	0.00	&	0.0            &       0.00            &       1.60    &       17.30   &       9307.87 &       0.00            \\
2.00	&	0.00	&	2.0$\times 10^{-5}$    &       8.76$\times 10^{-9}$    &       1.60    &       17.30   &       9307.87 &       1.35$\times 10^{-10}$   \\
2.00	&	0.00	&	2.0$\times 10^{-4}$    &       8.76$\times 10^{-8}$    &       1.60    &       17.30   &       9307.87 &       1.35$\times 10^{-8}$    \\
2.00	&	0.00	&	2.0$\times 10^{-3}$    &       8.76$\times 10^{-7}$    &       1.60    &       17.30   &       9307.86 &       1.35$\times 10^{-6}$    \\
2.00	&	0.00	&	2.0$\times 10^{-2}$    &       8.76$\times 10^{-6}$    &       1.60    &       17.30   &       9307.18 &       1.35$\times 10^{-4}$    \\
2.00	&	0.00	&	2.0$\times 10^{-1}$    &       8.61$\times 10^{-5}$    &       1.62    &       17.20   &       9240.81 &       1.39$\times 10^{-2}$    \\
\hline
\end{tabular}
\end{center}
\end{minipage}
\end{table*}

\section{Pulsation mode calculations}

\subsection{Perturbative analysis}

In order to calculate perturbative frequencies we follow closely the method of
\citet{Gough1990}. The perturbation equation, to order $\Omega^2$, can be written:
\begin{equation}
\label{eqn:LMN}
\mathscr{L}\vect{\xi} + \rho_{0} \omega^{2}\vect{\xi} = \omega
\mathscr{M} \vect{\xi} + \mathscr{N}\vect{\xi}
\end{equation}
where,
\begin{eqnarray}
\nonumber
\mathscr{L} \vect{\xi}&=& -\vect{\nabla} \left[ (p_{0} - \rho_{0} c_{0}^{2}) \vect{\nabla} \cdot
\vect{\xi} - \vect{\xi} \cdot \vect{\nabla} p_{0} \right] + p_{0}\vect{\nabla}(\vect{\nabla} \cdot
\vect{\xi}) \\
\nonumber
&&- \vect{\xi} \cdot \vect{\nabla}(\ln \rho_{0})\nabla p_{0}, \\ 
\mathscr{M}\vec{\xi}&=& -2i\rho_{0} \vect{v} \cdot \vect{\nabla} \vect{\xi}, \\ 
\nonumber
\mathscr{N}\vec{\xi}&=& \rho_{0} \left[-\vec{\xi}\cdot \vect{\nabla}(\vect{v}\cdot \vect{\nabla}
\vect{v}) + (\vect{v}\cdot\vect{\nabla})^{2}\vec{\xi}\right].
\end{eqnarray}

Here $p_{0}$, $\rho_{0}$ and $c_{0}$ are the equilibrium pressure, density and
sound speed respectively, $\omega$ is the angular pulsation frequency and
$\vect{v}_0$ the velocity field resulting from rotation.  $\vect{\xi}$ is the
Lagrangian displacement and is defined such that 
\begin{equation}
\vect{\xi}_{n{\ell}m} = \left(\xi(r) \Ylm(\theta,\phi),\eta(r) \frac{\partial \Ylm(\theta,\phi)}{\partial\theta},\frac{\eta(r)}{\sin{\theta}}\frac{\partial \Ylm(\theta,\phi)}{\partial \phi}\right)
\end{equation}
where $\xi$, $\eta$ are the radial and horizontal amplitude functions
respectively and $\Ylm$ is the spherical harmonic for a mode of degree $\ell$,
radial order $n$ and azimuthal order $m$. We choose to normalise $Y_{l}^{m}$
such that
\begin{equation}
\int_{0}^{2\pi}\int_{0}^{\pi} |\Ylm(\theta,\phi)|^2 \sin{\theta} \mathrm{d} \theta \mathrm{d} \phi = 1,
\end{equation}
and we normalise $\xi(r)$ such that $\xi(R)=1$.

The structural changes of the equilibrium model resulting from rotation cause
perturbations to the eigenfunction $\vect{\xi}$ and to the function
$\mathscr{L}\vect{\xi}$, denoted respectively by $\vect{\xi}_{1}$ and
$\mathscr{L}_{\Omega}\vect{\xi}$.  For further details we refer the reader to
\citet{Gough1990}. These perturbation quantities are substituted into equation
(\ref{eqn:LMN}) and we linearise the resulting expression to second order in
rotational velocity $\Omega$. This leads to the following expression for the
oscillation frequency, valid to second order in $\Omega$:
\begin{eqnarray}
\label{eqn:freq}
\nonumber
\omega_{n{\ell}m} &=&\omega_{0} + \omega_{\Omega 1} +
(2I\omega_{0})^{-1}\langle
\vect{\xi}^{*}_{0}\cdot(\mathscr{N}_{0} - \mathscr{L}_{\Omega} -
\rho_{\Omega}\omega^{2}_{0})\vect{\xi}_{0}\rangle \\
\nonumber && -
\omega^{2}_{\Omega 1}(2\omega_{0})^{-1}- \omega_{\Omega
1}I^{-1}\langle\rho_{0}\vect{\xi}^{*}_{0} \cdot\vect{\xi}_{1}\rangle  \\
\nonumber &&+
(2I)^{-1}\langle\vect{\xi}^{*}_{0}\cdot\mathscr{M}_{0}\vect{\xi}_{1}\rangle\\
&&+ \omega_{\Omega
1}(2I\omega_{0})^{-1}\langle\vect{\xi}^{*}_{0}\cdot\mathscr{M}_{0}\vect{\xi}_{0}\rangle,
\end{eqnarray}
\noindent where $\langle ... \rangle=\int_{0}^{2\pi}\int_{0}^{\pi}\int_{0}^{R}
... x^{2}\sin\theta \rm{d} x\rm{d}\theta\rm{d}\phi $ is, to first order,
the volume integral over the interior of the star. $\omega_{0}$ is the
unperturbed oscillation frequency, 
\begin{equation}
\label{eqn:I}
I=\int_{0}^{R}\rho_{0}x^{2}\left[\xi^{2}+\ell(\ell+1)\eta^{2}\right]\rm{d}x,
\end{equation}
\noindent and $\omega_{\Omega 1}$ contains the terms to first order in $\Omega$ and can be expressed as:
\begin{equation}
\label{eqn:omega1}
\omega_{\Omega 1}=\frac{m}{I}\int_{0}^{R}\Omega\rho_{0}x^{2}\left[(\xi-\eta)^{2}+(\ell(\ell+1)-2)\eta^{2}\right]\rm{d}x.
\end{equation}

The ADIPLS adiabatic pulsation package \citep{adipls} was then modified
to include second order rotational effects, and subsequently used to
calculate perturbative frequencies.

\subsection{Two-dimensional calculations}

In order to carry out the two-dimensional pulsation calculations, we used the
same approach as in \citet{Reese2009}.  The pulsation equations are expressed in
terms of a new coordinate system which follows the shape of the star, then
projected onto the spherical harmonic basis, and finally solved using the
pulsation code TOP \citep[Two-dimensional Oscillation Program,][]{Reese2009}.

The new coordinate system can be represented by $(\zeta,\theta,\phi)$ where
$\zeta$ is the radial coordinate, $\theta$ the colatitude, and $\phi$ the
longitude.  The relationship between $\zeta$ and $r$, the distance from the
origin, is given by the following two equations:
\begin{eqnarray}
\label{eq:domain1}
r(\zeta,\theta) &=& (1-\epsilon)\zeta+\frac{5\zeta^3-3\zeta^5}{2}
                 \left( R_s^\star(\theta) - 1 + \epsilon \right), \\
\label{eq:domain2}
r(\zeta,\theta) &=& 2\epsilon + (1-\epsilon) \zeta \nonumber \\
                & & + \left( 2\zeta^3 - 9\zeta^2+12\zeta-4\right)
                   \left( R_s^\star(\theta) - 1 - \epsilon \right),
\end{eqnarray}
where $\epsilon = 1 - \frac{\Rps}{R}$, $R_s^{\star}(\theta) = \frac{\Rs}{R} -
P_2(\cost) \frac{\Rs h_{\Omega}(\Rs)}{R}$, $\Rps$ is the polar radius at the
last grid point and $\Rs$ the radial coordinate at the last grid point. The
first equation applies to the first domain, \ie for $\zeta \in [0,1]$, which
corresponds to the star.  It must be noted that contours of constant $x$ and
$\zeta$ values do not in general coincide, except at the last grid point where
$x = \Rs$ and $\zeta = 1$, because only the former corresponds to isobars.  As a
result, the model must be interpolated onto this new grid.  The second equation
applies to the second domain, \ie for $\zeta \in [1, 2]$, which lies outside the
star and in which only the perturbation to the gravitational potential is used. 
This coordinate system ensures that the standard regularity conditions can be
used in the centre since it behaves like the spherical coordinate system in the
centre, and it allows the use of simple boundary conditions both on the stellar
surface for the Lagrangian displacement and pressure fluctuations, and
on the outer boundary of the second domain for the perturbation to the
gravitational potential.

The pulsation equations are expressed in terms of the Lagrangian displacement,
$\vect{\xi}$:
\begin{eqnarray}
\label{eq:continuity}
0 &=& \rho + \div \left(\rho_0 \vect{\xi} \right), \\
\label{eq:Lagrange}
0 &=&  \vlp^2 \rho_0 \vect{\xi} - 2i\vlp \rho_0 \vect{\Omega} \times \vect{\xi} \nonumber \\
  & & -\grad p + \rho \vect{g}_\mathrm{eff} - \rho_0 \grad \Psi, \\
\label{eq:adiabatic}
0 &=& p + \vect{\xi}\cdot \grad p_0 - c_0^2
      \left(\rho + \vect{\xi} \cdot \grad \rho_0 \right), \\
\label{eq:Poisson}
0 &=& \lapl \Psi - \Lambda \rho,
\end{eqnarray}
where $\rho$, $p$ and $\Psi$ are the density, pressure and gravitational
potential fluctuations.  This is the same set of equations as what is given in
the appendix of \citet{Reese2009} except that there is no gradient of the
rotation rate, since the rotation profile is uniform, and each occurrence of
$\omega$ has been replaced by $-\omega$, so as to ensure that prograde modes
correspond to positive values of $m$, the azimuthal order.  Furthermore, in
order to increase the accuracy of the calculations, the effective gravity has
been calculated from the gradient of the total potential rather than from the
gradient of the pressure divided by the density.  For explicit expressions in
terms of the spheroidal coordinate system, we refer the reader to
\citet{Reese2009}.

Calculating pulsation modes in models based on the ASTEC code \citep{astec} is
similar to calculating modes in models based on the Self-Consistent Field (SCF)
method \citep{MacGregor2007}, although there are some noteworthy differences. 
ASTEC models include atmospheres which means that the associated radial grids
become very dense near the surface.  As a result, care must be taken when
expressing the radial differential operator in algebraic form.  For instance,
using the fourth order finite difference implemented in \citet{Reese2009} leads to
numerically unstable behaviour.  This has been replaced with a more stable
version of fourth order finite differences, in which the differential equations
are solved on carefully chosen intermediate grid points.

\section{Validity domains for perturbative calculations}
\label{sect:comparison}

\begin{figure}
  \includegraphics[width=\columnwidth]{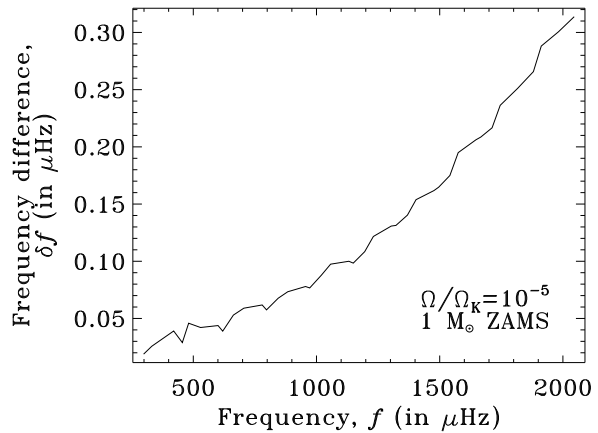}
  \caption{A comparison of $m=0$ modes for a 1 $M_{\odot}$ model rotating at $\Omega =
           10^{-5}\Omega_\mathrm{K}$.  The radial orders are $n=1,10$ and the
           harmonic degrees $\ell=0,3$.  These differences are larger than 0.08
           $\mu$Hz, the error bar on a CoRoT long run, and must therefore be
           taken into account when constructing validity domains.}
  \label{fig:df}
\end{figure}

Before establishing validity domains for second order perturbative methods, it
is important to compare the two methods for non-rotating (or nearly
non-rotating) stars.  Differences arise from the different formulations of the
pulsation equations and should be correctly characterised before proceeding to
establish validity domains.  Figure~\ref{fig:df} shows the differences between
$m=0$ modes calculated with the two methods.  Although this error is small, it
is larger than the error bar from a CoRoT long run and should therefore be
taken into account when constructing validity domains.

\begin{figure}
  \includegraphics[width=\columnwidth]{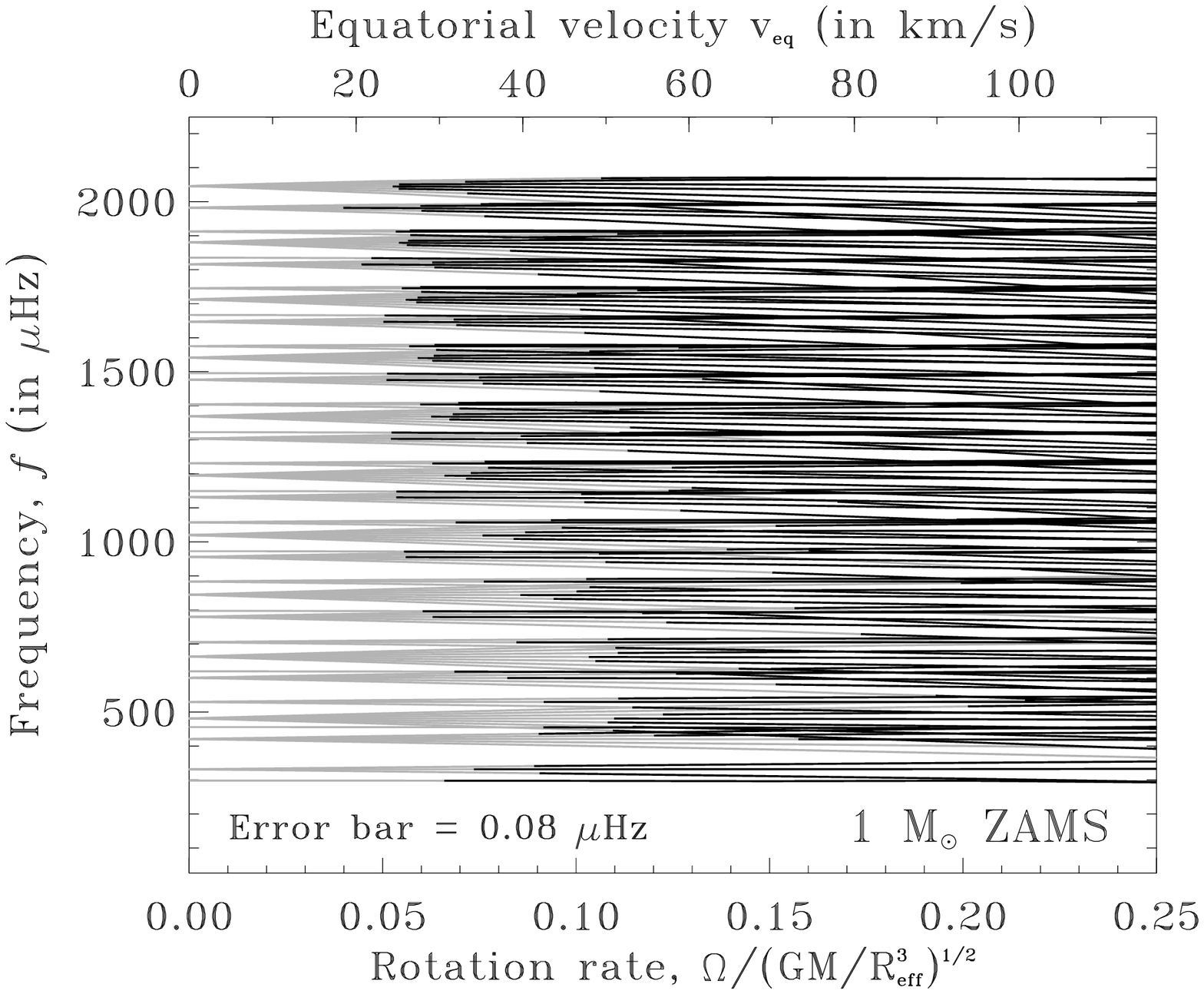} \\
  \includegraphics[width=\columnwidth]{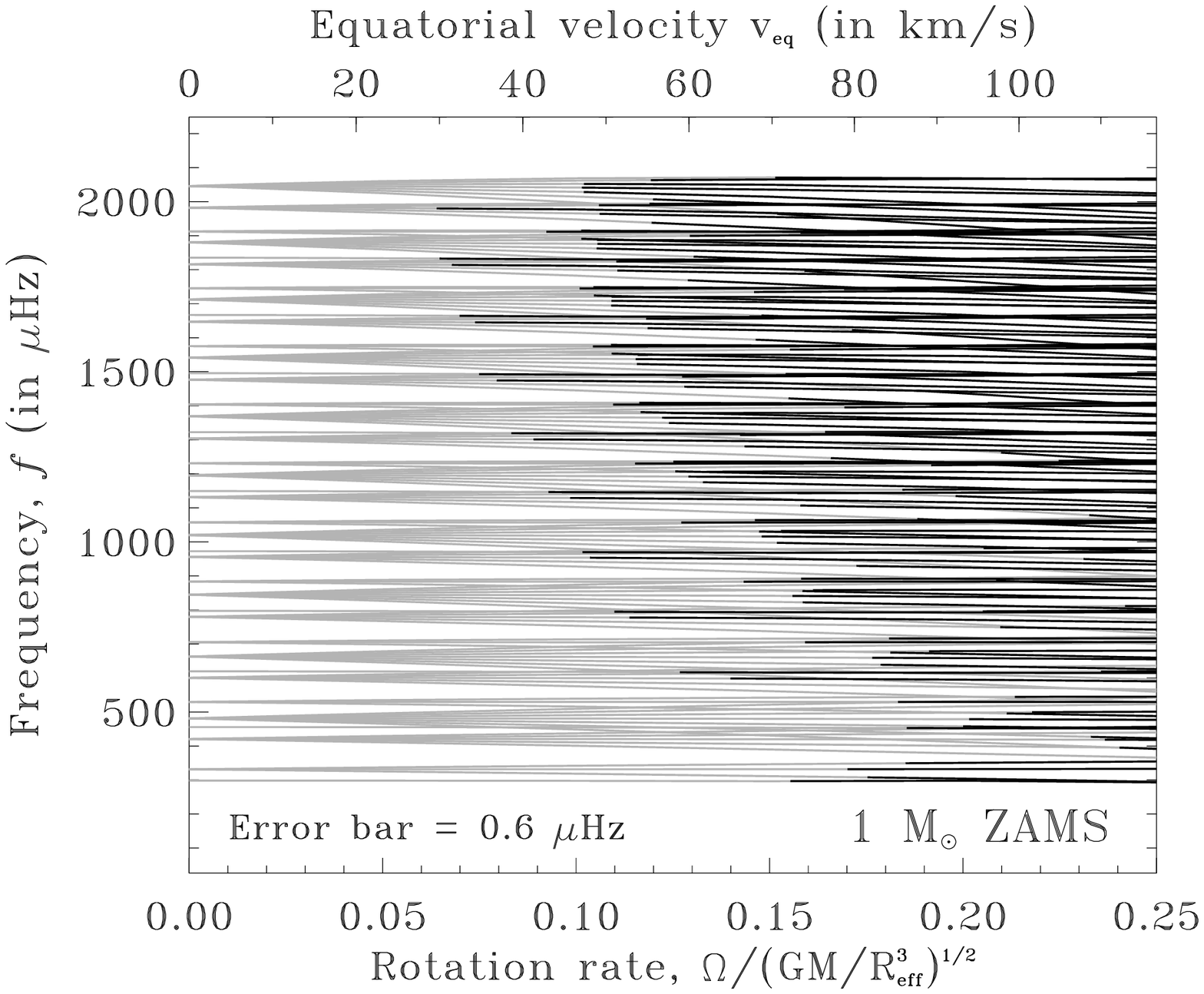} \\
  \includegraphics[width=\columnwidth]{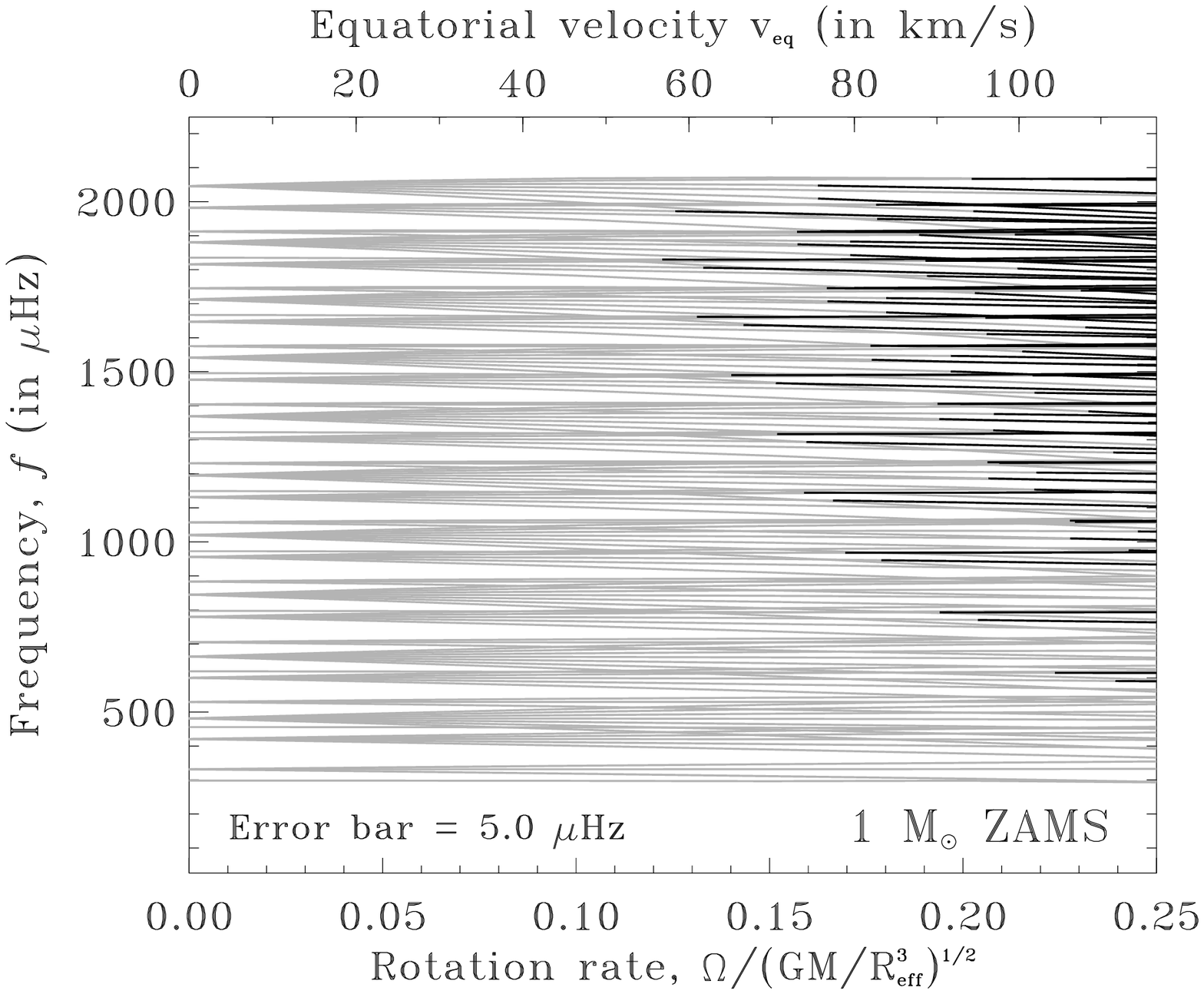}
  \caption{Validity domains for second order perturbative frequencies of 1
           $M_{\odot}$ ZAMS models for three different error bars, the first two
           corresponding to CoRoT error bars.  These figures are analogous to
           Fig.~4 of \citet{Reese2006}, which were established for an $N=3$
           polytropic model.}
  \label{fig:comp}
\end{figure}

Figure~\ref{fig:comp} shows a comparison between the two methods for three
different error bars.  The first two correspond to a long and short CoRoT run
and the third corresponds to 2.3 days of observation.  The radial orders of the
modes are $n=1-10$, the harmonic degrees $\ell=0-3$ and the azimuthal orders
$m=-\l$ to $\l$.  Frequency multiplets calculated using perturbative theory were
shifted using the differences which are plotted in Fig.~\ref{fig:df}.  This
leads to somewhat larger and more realistic validity domains, especially for the
smallest error bar. Based on these validity domains, non-perturbative effects
will start to play an important role beyond $\veq = 30\,\,\kms$ for a long run,
and $\veq = 50\,\,\kms$ for a short run, in a $1\,\,M_{\odot}$ star.  However,
given that solar-like pulsators tend to oscillate with radial orders between
$15$ and $25$, these limits are expected to be lower, as based on the trends
which can be seen in Fig.~\ref{fig:comp}.

\begin{figure}
  \includegraphics[width=\columnwidth]{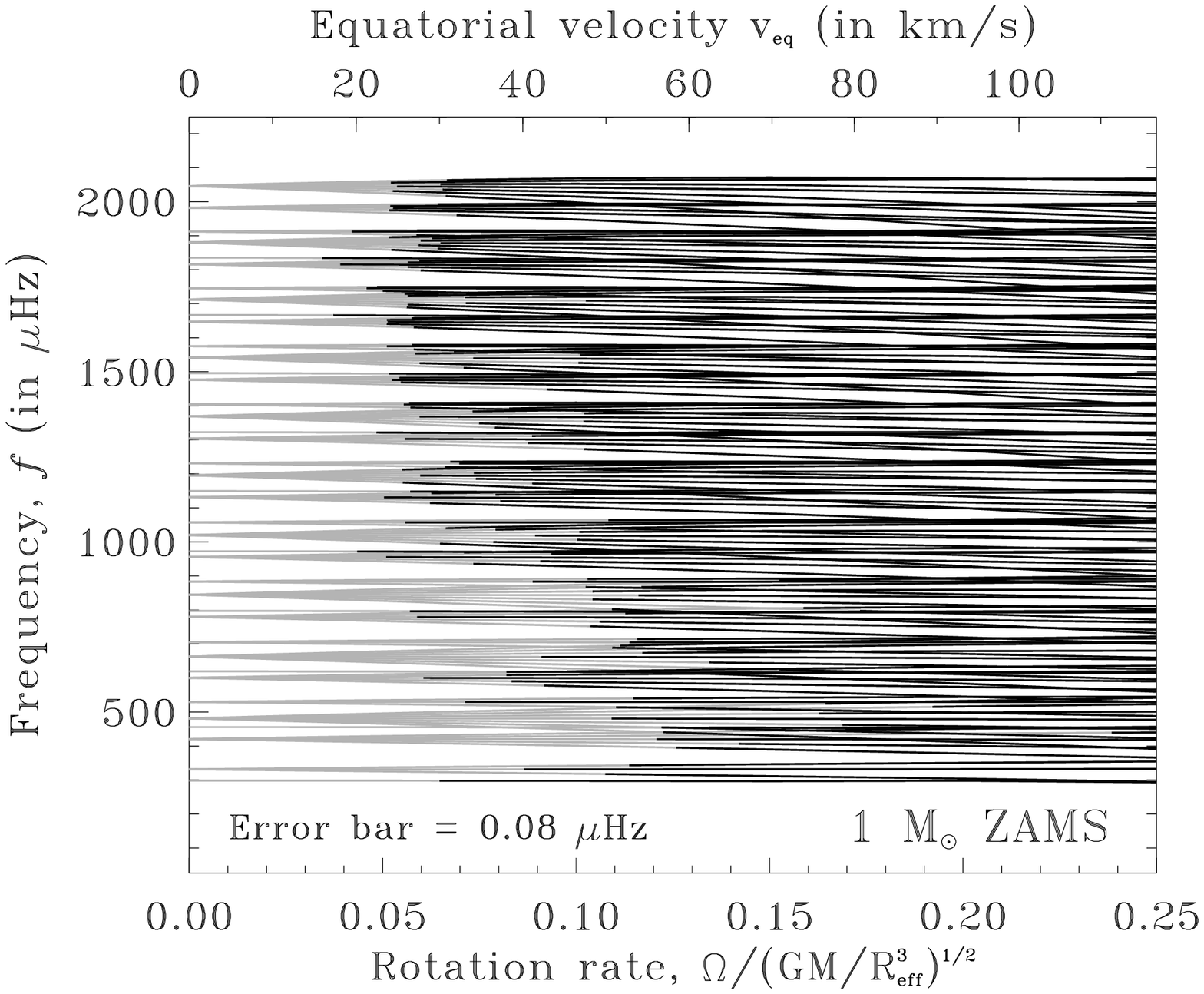} \\
  \includegraphics[width=\columnwidth]{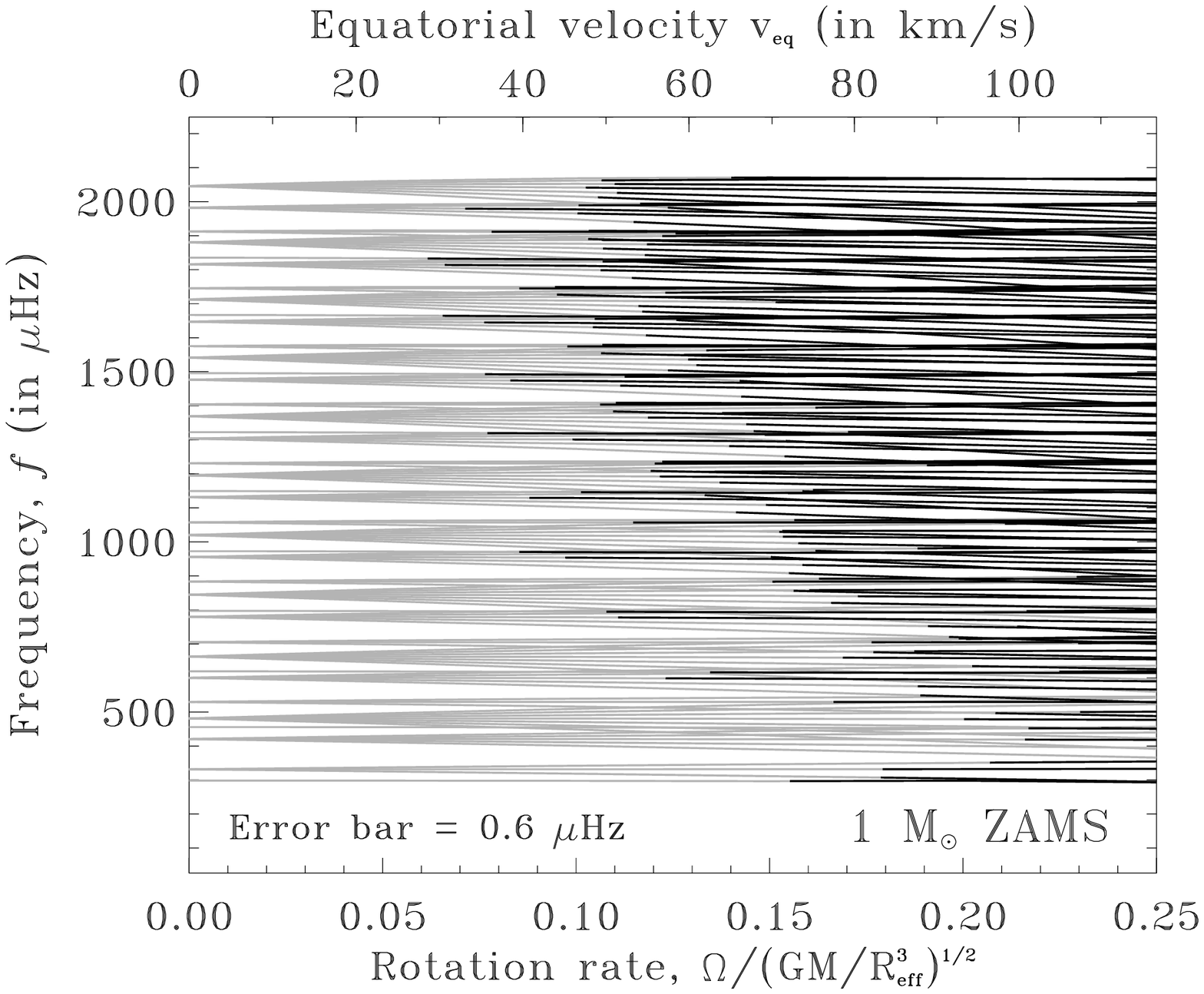} \\
  \includegraphics[width=\columnwidth]{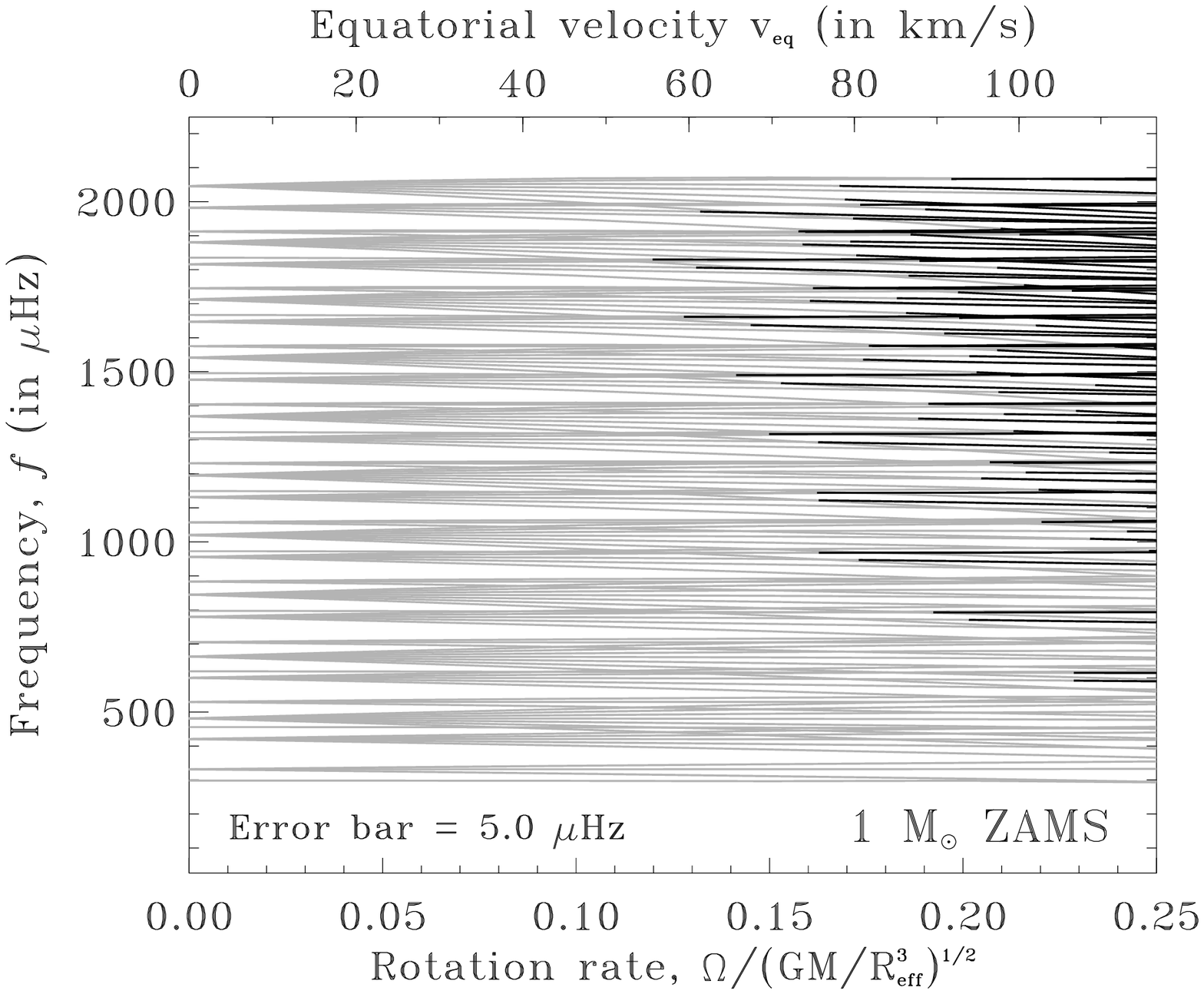}
  \caption{Same as Fig.~\ref{fig:comp} except that the perturbative frequencies
           have been replaced with second order polynomial fits to the
           non-perturbative frequencies.}
  \label{fig:comp_fit}
\end{figure}

If the perturbative frequencies are replaced with polynomial fits to the
non-perturbative frequencies, similar validity domains are obtained, as can be
seen in Fig.~\ref{fig:comp_fit}.  This comparison shows that there is a good
agreement between the polynomial coefficients and the coefficients deduced from
perturbative theory, thus providing a further check that both the perturbative
and two-dimensional approach agree at low rotation rates.

\begin{figure}
  \includegraphics[width=\columnwidth]{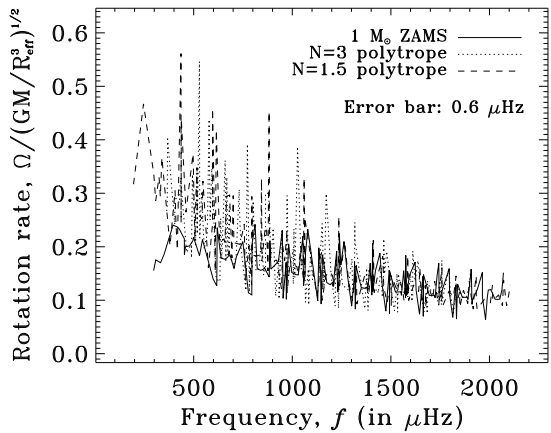}
  \caption{Comparison between between the validity domains for two polytropic
           models and the 1 $M_{\odot}$ ZAMS model, for the $0.6\,\,\mu$Hz error
           bar.  The threshold separating the regions where perturbative methods
           are valid and not valid is plotted so as to allow the superposition
           of the different domains.}
  \label{fig:threshold}
\end{figure}

It is then interesting to compare these validity domains with those obtained
from polytropic models.  Figure~\ref{fig:threshold} shows such a comparison. It
includes the validity domain for $N=3$ and $N=1.5$ polytropic models and
compares them with that of the 1 $M_{\odot}$ ZAMS model using the $0.6\,\,\mu$Hz
error bar.  The polytropic models are scaled so as to have the same mass and
approximately the same radii\footnote{The polar and equatorial radii were made
to satisfy $\frac{\Rp+\Req}{2} = 0.89\,\,R_{\odot}$, which is approximately true
of the 1 $M_{\odot}$ ZAMS model.} as the 1 $M_{\odot}$ ZAMS model.  In order to
superimpose the different validity domains, we have simply plotted the threshold
between the region where the perturbative approach is valid, and the region
where a two-dimensional calculation becomes necessary. As can be seen from the
figure, the validity domains are quite similar.  At high frequencies, they follow the same tendency
as a function of frequency and have a similar spread around the mean value.  At
low frequencies, the threshold for the 1 $M_{\odot}$ model seems to be, on
average, lower than that of the polytropic models.  This effect is, however,
likely to be an artifact due to the cutoff at $\Omega =
0.25\,\,\Omega_{\mathrm{K}}$ for the $1\,\,M_{\odot}$ frequency calculations
(see Fig.~\ref{fig:comp}).

\section{Close frequency pairs}
\label{sect:frequency_pairs}

In \citet{Espinosa2004}, it was pointed out that adjacent frequencies within a
multiplet tend to pair up and come close together, except for the most
retrograde mode (\textit{i.e.} $\l = -|m|$ sectoral mode).  It was then suggested
that this phenomena could explain the close frequency pairs observed in
\citet{Breger2002} (see also \citealt{Breger2006} and \citealt{Breger2006b}).
A similar pairing up of modes also occurs for the frequencies in \citet{Reese2006}
as well as the non-perturbative frequencies presented here.
In what follows, we will consider 2 $M_{\odot}$ ZAMS models, as this is more representative
of $\delta$ Scuti stars.  The second half of Table~\ref{tab:models} gives the characteristics
for a selection of these models.

\begin{figure*}
  \includegraphics[width=\columnwidth]{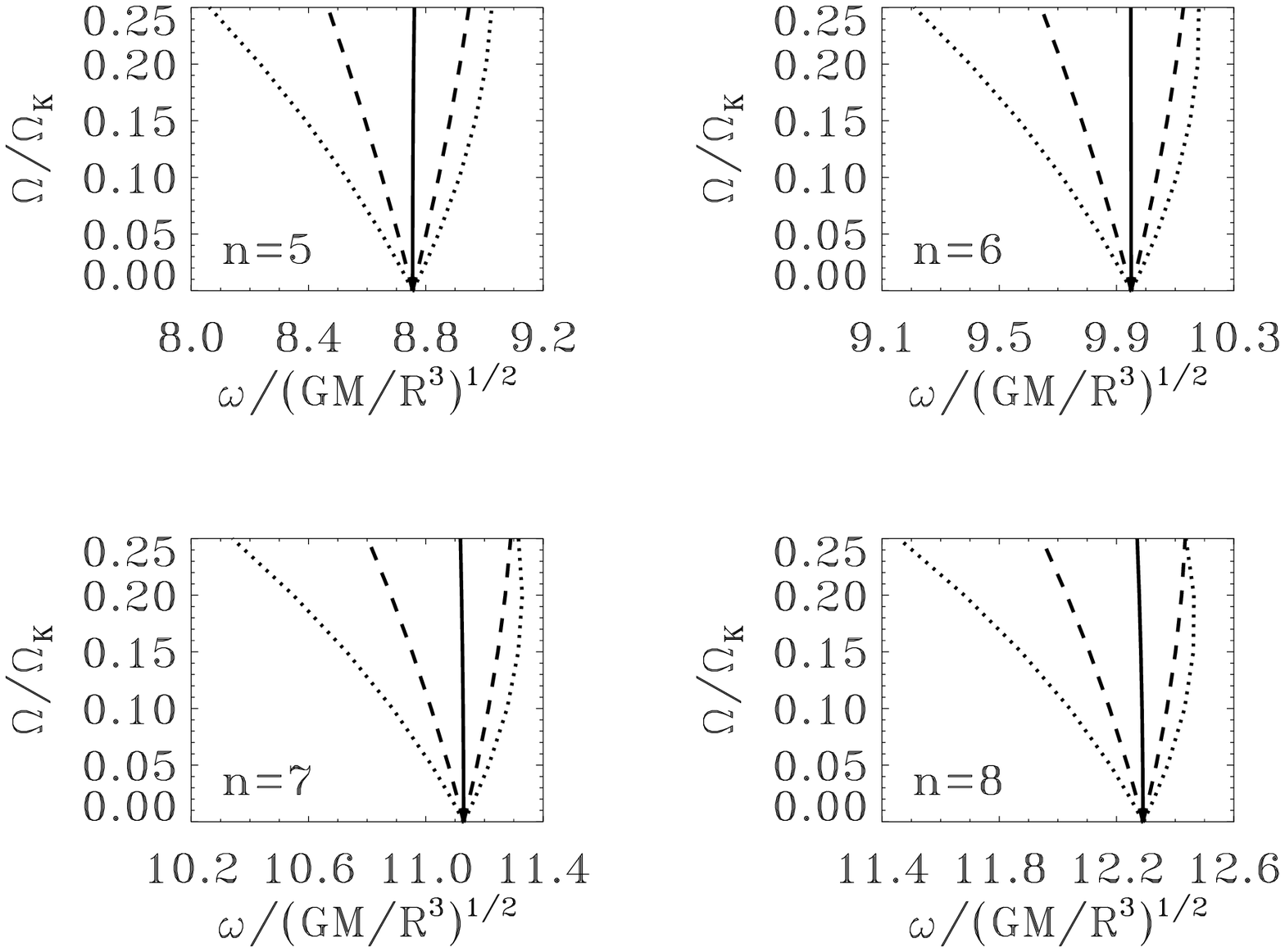} \hfill
  \includegraphics[width=\columnwidth]{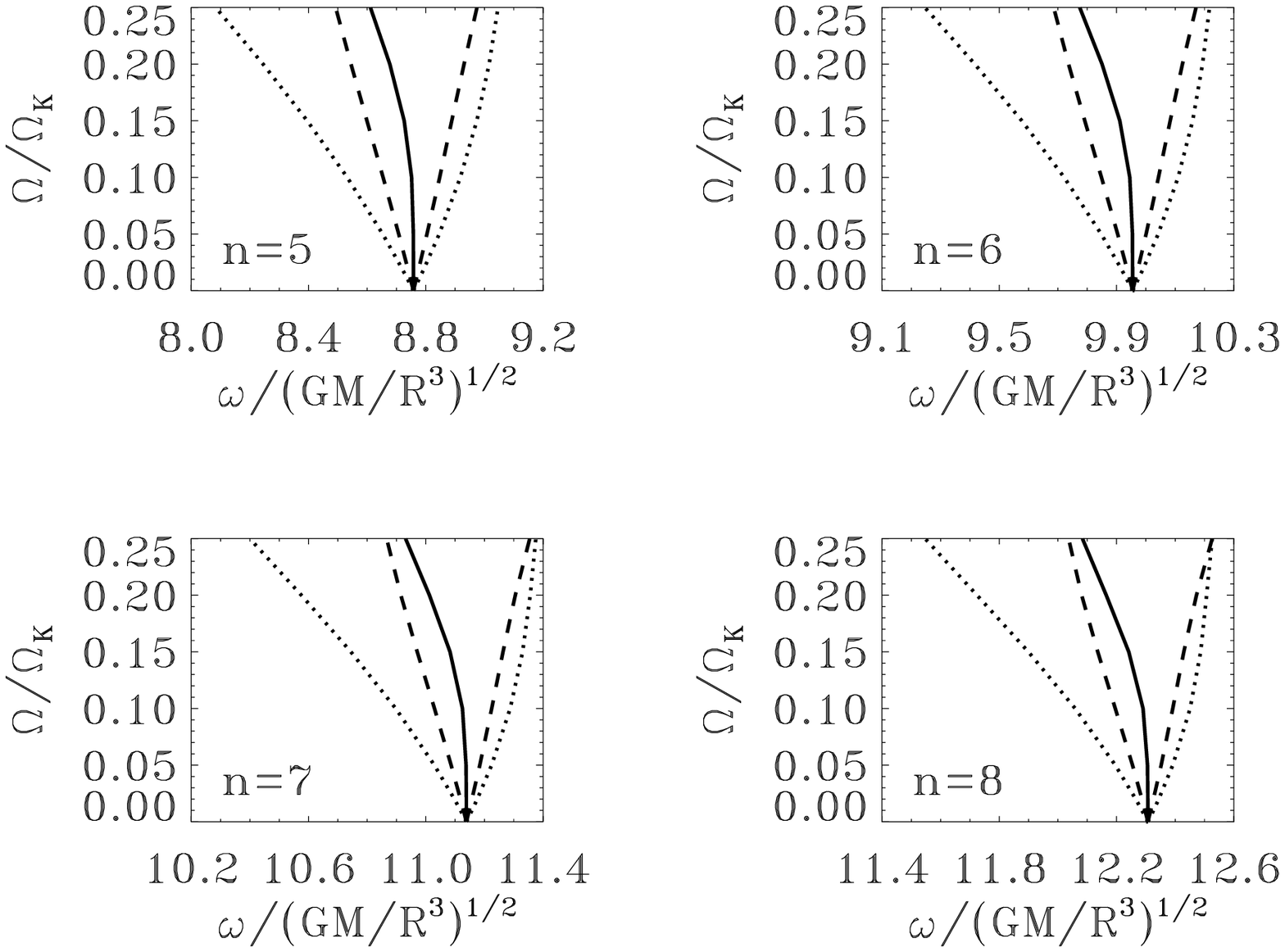} \\
  \caption{Four $\ell=2$ multiplets calculated using perturbative \textit{(left)}
           and two-dimensional \textit{(right)} calculations in 2 $M_{\odot}$ models.  Frequencies only
           pair up in the two-dimensional calculations.  In the perturbative
           case, the spacings decrease uniformly when going from the most
           retrograde to the most prograde mode.}
  \label{fig:pairs}
\end{figure*}

Figure~\ref{fig:pairs} shows four sets of $\l=2$ multiplets calculated using
both perturbative and two-dimensional calculations.  As can be seen in the
figures, only the two-dimensional calculations lead to this behaviour.  The
perturbative calculations produce, instead, spacings which decrease uniformly
when going from the most retrograde to the most prograde mode.  However,
the effects of avoided crossings have not been included in the perturbative
calculations, so it remains to be seen whether including this effect can
also produce a pairing up of adjacent frequencies.

\begin{figure}
  \includegraphics[width=\columnwidth]{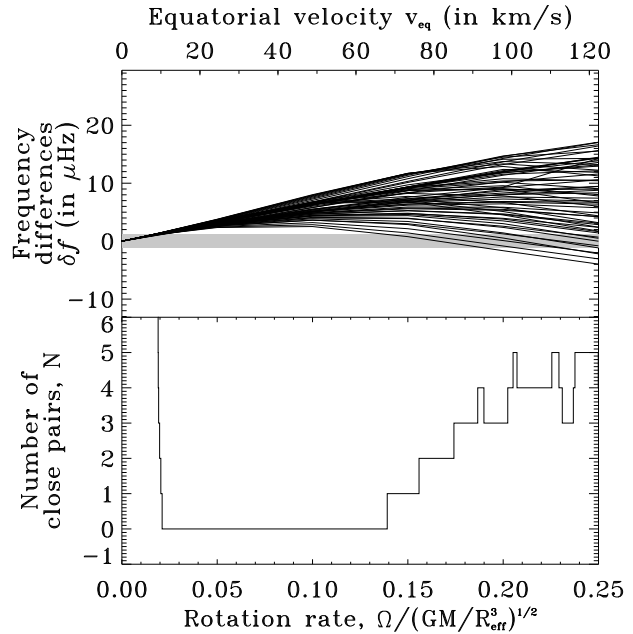}
  \caption{\textit{Upper panel}: frequency differences between adjacent modes
  in the $\l=1-3$ frequency multiplets of 2 $M_{\odot}$ models, as a function of the rotation rate.
  The frequency span of the spectrum is approximately $250-1060\,\,\mu$Hz.  The grey band
  in the middle corresponds to differences which are less than 0.1 c/d.
  \textit{Lower panel}: number of frequency differences below 0.1 c/d as a
  function of the rotation rate.  This number is generally quite low except
  for slow rotation rates, where the average spacing between frequencies in a
  multiplet is comparable to 0.1 c/d.}
  \label{fig:pairs_count}
\end{figure}

It is then interesting to investigate whether this effect could explain the
close frequency pairs observed in \citet{Breger2006}.  In
Fig.~\ref{fig:pairs_count}, we compare the frequency differences of mode pairs
to a frequency separation of 0.1 c/d (\textit{i.e.} $1.1574\,\,\mu$Hz), a
typical threshold for observed frequency pairs. As can be seen from the upper
panel, these differences remain larger on average than 0.1 c/d.  The lower panel
shows the number of sufficiently close frequencies as a function of the rotation
rate.  As can be seen, this number is not very large for any given rotation
rate, except when the rotation rate is around or below 0.1 c/d.  However, 18
close frequency pairs were observed in FG Vir \citep{Breger2006b}, and it's
equatorial velocity is estimated to be $66\pm16\,\,\kms$, as based on the
modelling of line profile variations \citep{Zima2006}.  As a result, the pairing
up of adjacent modes in frequency multiplets do not seem to account for close
frequency pairs in $\delta$ Scuti stars.

\begin{figure}
  \includegraphics[width=\columnwidth]{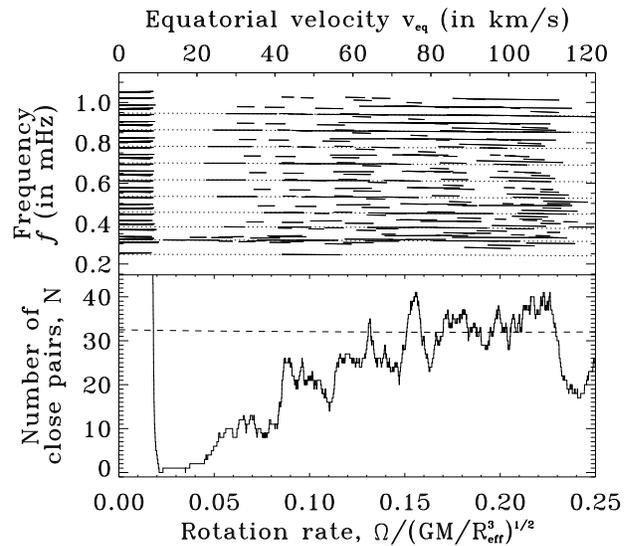}
  \caption{\textit{Upper panel}: close frequency pairs in 2 $M_{\odot}$ models as a function of the
  rotation rate for the frequency spectrum ($n=1-10$, $\l=0-3$, $m=-\l$
  to $\l$).  The horizontal dotted lines show the location of radial
  modes.  \textit{Lower panel}: number of close frequency pairs as a function of
  the rotation rate.  The dashed line shows the number of close frequency pairs
  that is expected for a Poisson distribution.}
  \label{fig:all_pairs}
\end{figure}

Of course, it is always possible to look at all of the frequency differences for
a given spectrum.  This approach has been used by \citet*{Lenz2008} to study 3
$\delta$ Scuti stars, including FG Vir.  Here, we will push the analysis slightly
further by looking at how the number of close frequency pairs depends on the
rotation rate.  Figure~\ref{fig:all_pairs} shows the number of close frequency
pairs as a function of the rotation rate, as well as where they occur. The
dashed line in the lower panel gives the expected number of pairs,
$N_{\mathrm{exp.}}$, as based on a Poisson distribution:
\begin{equation}
  N_{\mathrm{exp.}} = \left(N_{\mathrm{mode}}-1\right) \left[1-
  \exp\left(-\frac{\left(N_{\mathrm{mode}}-1\right) \delta f}{\Delta f}\right)\right]
\label{eq:Nexp}
\end{equation}
where $N_{\mathrm{mode}}$ is the number of modes in the spectrum, $\Delta f$ the
frequency span of the spectrum, and $\delta f$ the target frequency separation
(\textit{i.e.} 0.1 c/d).  At a sufficient rotation rate, the average number of
close frequency pairs matches $N_{\mathrm{exp.}}$ indicating that the spectrum is
behaving like a Poisson distribution.  As was pointed out by
\citet{Lignieres2008} and \citet{Lignieres2009}, the frequency spectrum of
rapidly rotating stars is subdivided into classes of regular modes, such as
these low degree modes, and chaotic modes, the frequencies of which follow
Poisson and Wigner distributions respectively, provided there are no selection
effects.


The number of close frequency pairs is much higher than in the previous case and
higher than the number of pairs observed in FG Vir.  However,
Fig.~\ref{fig:all_pairs} is based on 160 frequencies, whereas there are
67 independent frequencies detected in FG Vir. Applying Eq.~(\ref{eq:Nexp})
to FG Vir's frequency spectrum yields 10.4 close frequency pairs, which is slightly over
half the number of observed close frequency pairs in this star. Also, as pointed out in \citet{Lenz2008},
there are 7 frequency pairs where the separation is below 0.01 c/d.  Equation~(\ref{eq:Nexp})
would yield 1.12 very close frequency pairs for FG Vir. Furthermore, according to \citet{Breger2006b},
close frequency pairs seem to cluster around radial modes, whereas a number of
pairs in Fig.~\ref{fig:all_pairs} are close to the midpoints between consecutive
radial modes.  Hence, an astrophysical origin is still needed to account for
this phenomena.

An interesting feature appears in Fig.~\ref{fig:all_pairs}.  Between
approximately $0.02\,\,\Omega_{\mathrm{K}}$ and $0.05\,\,\Omega_{\mathrm{K}}$,
few frequency pairs are detected.  A simple explanation is that the
rotation rate is sufficient to keep individual multiplet components far enough
apart, but too small to cause frequency multiplets to overlap.  Of course,
including modes with higher $\ell$ values would probably reduce this gap by
introducing new multiplets into the frequency spectrum.  However, as was pointed
out by \citet{Lignieres2006} and \citet{Lignieres2009}, cancellation effects for
such modes are more effective at lower rotation rates, \textit{i.e.} where close
frequencies pairs are lacking.  Table~\ref{tab:stars_with_pairs} gives a list of
$\delta$ Scuti stars where close frequencies have been reported.  Interestingly,
none of the stars fall in this gap.  Of course, a larger number of stars would
need to be analysed to see whether this gap remains or whether it is simply due
to poor statistics. Furthermore, some of the stars in
Table~\ref{tab:stars_with_pairs}, such as BI CMi \citep{Breger2002b}, BV Cir
\citep*{Mantegazza2001} and 44 Tau \citep{Zima2007}, are evolved.  Consequently,
their pulsation spectra are likely to contain acoustic, gravity (or
gravito-inertial) and possibly mixed modes.

\begin{table}
\begin{tabular}{lrrl}
\hline
\hline
\multicolumn{1}{c}{\textbf{Star}} &
\multicolumn{1}{c}{\textbf{HD}} &
\multicolumn{1}{c}{$\mathbf{\vsini}$} &
\multicolumn{1}{c}{\textbf{Reference}} \\
&
\multicolumn{1}{c}{\textbf{number}} &
\multicolumn{1}{c}{\textbf{($\mathbf{\kms}$)}} &
\\
\hline
BI CMi   &  66853 & $76\pm1$       & \citet{Breger2002b}    \\
XX Pyx   &        & $52\pm2$       & \citet{Handler1997}    \\
V509 Per &  18878 & $134$          & \citet{Bush2008}       \\
4 CVn    & 107904 & $112\pm3^a$    & \citet{Bush2008}       \\
FG Vir   & 106384 & $21.6\pm0.3^b$ & \citet{Zima2006}       \\
BV Cir   & 132209 & $96.5\pm1$     & \citet{Mantegazza2001} \\
BW Cnc   &  73798 & $200\pm11$     & \citet{Fossati2008}    \\
44 Tau   &  26322 & $2\pm1^c$      & \citet{Zima2007}       \\
\hline
\end{tabular} \\
\caption{\normalsize Projected equatorial velocities of $\delta$ Scuti stars
with observed close frequency pairs (this list is based on \citealt{Breger2002}
and \citealt{Lenz2008}). The first two columns identify the star.  The third and
fourth columns give the projected equatorial velocity and the reference where
this velocity is obtained.  \small $^a$ \citet{Bush2008} give a list of several
values.  The value given here is the average plus or minus the standard
deviation. $^b$ Although the projected equatorial velocity is small,
\citet{Zima2006} estimate that the true equatorial velocity is
$66\pm16\,\,\kms$. $^c$ \citet{Zima2007} estimate that the true equatorial
velocity is $3\pm2\,\,\kms$.}
\label{tab:stars_with_pairs}
\end{table}

\section{Conclusion}

In this study, the effects of stellar rotation on acoustic pulsation frequencies
have been investigated using a second order perturbative and a two-dimensional
approach.  A comparison of the two shows that perturbative methods are valid for
equatorial velocities up to $30$ and $50\,\,\kms$ for a CoRoT long and short
run, respectively.  The associated validity domains closely match the domains
previously obtained in \citet{Reese2006} for polytropic models. These results
show that perturbative methods, which are simpler to work with and less
demanding numerically, will remain useful for a number of stars.  Nonetheless,
various qualitative and quantitative differences appear between this approach
and two-dimensional calculations at sufficient rotation rates.  Among these is
the pairing up of adjacent modes in frequency multiplets.  This phenomena, first
noted in \citet{Espinosa2004}, also shows up in the two-dimensional calculations
presented here but not in the second order perturbative calculations, in which
the spacing between consecutive modes varies monotonically.  The cause of this
pairing up remains a mystery, given that different modes in a multiplet are not
coupled due to their differing azimuthal orders.

The case of observed close frequency pairs is then discussed.  In particular, it
is shown that adjacent mode pairs in frequency multiplets do not come close
enough together to provide a likely explanation, as opposed to what was
previously suggested \citep{Espinosa2004}.  A systematic search for all of the
close frequencies pairs in the calculated spectra was then carried out.  Results
showed that at sufficient rotation rates, the number of close pairs of
low degree acoustic modes matched what is expected based on a Poisson
distribution.  Even then, this does not match the number and distribution of
close pairs observed in stars like FG Vir thereby favouring an astrophysical
origin to the phenomena.  At lower rotation rates, where frequency multiplets do
not overlap, a lack of close frequency pairs is observed in the calculated
spectra, except at the lowest rotation rates where the rotational shift is
small.  Delta Scuti stars currently reported as having close frequency pairs do not
lie in this interval, but a more systematic study is required to see 
to what extent this is due to poor statistics, and to see what effects including
gravity type modes has for more evolved stars.

\section*{Acknowledgements}
The authors wish to thank the referee for useful comments which have
helped to clarify the manuscript and improve the scientific discussion.
DRR gratefully acknowledges support from the CNES (``Centre National d'Etudes
Spatiales'') through a postdoctoral fellowship, the UK Science and Technology
Facilities Council through grant ST/F501796/1, and the European Helio- and
Asteroseismology Network (HELAS),a major international collaboration funded by
the European Commission's Sixth Framework Programme.  This research has made use
of the SIMBAD database, operated at CDS, Strasbourg, France.

\bibliographystyle{mn2e}
\bibliography{biblio}
\label{lastpage}
\end{document}